\documentclass{ptapap}
\usepackage{color}
\usepackage[utf8]{inputenc}
\usepackage{graphicx}
\usepackage{indentfirst}
\usepackage{amssymb}
\usepackage{hyperref}
\hypersetup{colorlinks=true,citecolor=blue}

\author{Agnieszka Kurcz}[UJ,ZG]
\author{Magdalena Krupa}[UJ,ZG]
\author{Maciej Bilicki}[LU,ZG,UC]
\author{Aleksandra Solarz}[NC,ZG]
\author{Agnieszka Pollo}[UJ,NC,ZG]
\author{Katarzyna Małek}[NC,ZG]

\affil[UJ]{Astronomical Observatory, Jagiellonian University, 
\\ ul.\ Orla 171, 30-244 Kraków, Poland}
\affil[ZG]{Janusz Gil Institute of Astronomy, University of Zielona Góra, 
\\ ul.\ Lubuska 2, 65-265 Zielona Góra, Poland} 
\affil[LU]{Leiden Observatory, Leiden University, 
\\ P.O. Box 9513 NL-2300 RA  Leiden, The Netherlands}  
\affil[NC]{National Centre for Nuclear Research, Astrophysics Division,
\\ ul. Hoża 69, 00-681 Warszawa, Poland}
 \affil[UC]{Astrophysics, Cosmology and Gravity Centre, Department of Astronomy, \\University of Cape Town, 
 Rondebosch, South Africa}

\title{Automatised classification of WISE sources: first results, future prospects}

\begin{document}

\maketitle

\begin{abstract}

We present the first results of our dedicated programme of automatised classification of galaxies, stars and quasars in the mid-infrared all-sky data from the WISE survey. We employ the Support Vector Machines (SVM) algorithm, which defines a hyperplane separating different classes of sources in a multidimensional space of arbitrarily chosen parameters. This approach consists of four general steps: 1) selection of the training sample, 2) selection of the optimal parameter space, 3) training of the classifier, 4) application to target data. Here, as the training set, we use sources from a cross-correlation of the WISE catalogue with the SDSS spectroscopic sample. The performance of the SVM classifier was tested as a function of size of the training set, dimension of the parameter space, WISE apparent magnitude and Galactic extinction. We find that our classifier provides promising results already for three classification parameters: magnitude, colour and differential aperture magnitude. Completeness and purity levels as high as 95\% are obtained for quasars, while for galaxies and stars they vary between 80--95\% depending on the magnitude, deteriorating for fainter sources.

\end{abstract}

\section{Introduction}
Today's flood of astronomical data gathered in ever wider and deeper datasets comes at a price of inability to obtain spectra for the majority of the observed sources. Without knowing the spectral features, classification of objects in large photometric samples becomes far from straight-forward, especially at the faint end and in the low signal-to-noise regime. A good example is the all-sky catalogue from the Wide-field Infrared Survey Explorer (WISE, \citealt{WISE}), providing various pieces of photometric and astrometric information for almost $10^9$ sources, without however any object type identification. For such amount of data, human visual verification of sources is clearly infeasible except for some very small subsamples. Limited number of photometric bands and a low detection rate in some of them, together with sources overlapping in multi-colour space, make also traditional approaches to separate objects, such as through colour cuts, not always effective in this dataset.

Current approaches towards identifying specific source types in WISE consist mainly in cross-matching this dataset with an external one (e.g.\ SDSS) and calibrating some magnitude and colour cuts to be applied for a specific subsample preselection (e.g.\ \citealt{Stern12} for AGNs or \citealt{TuWa13} for AGB stars). Application of such cuts to the all-sky WISE data may however give biased results, for such reasons as non-representativeness of the calibration sample, variations in WISE source detection rate (cf.\ \citealt{Secrest15}) or blending and varying stellar populations depending on sky position, leading to variations is source effective colours (e.g.\ \citealt{Ferraro15}).

A possible way to avoid these issues is to rely on automatised classification through machine learning (ML) algorithms, such as Support Vector Machines (SVM). An example application using WISE data is provided in \cite{KoSz15}, where however the depth of the resulting galaxy catalogue is limited by much shallower  2MASS. Our aim is to go beyond such limitations and apply ML to the majority of WISE sources, ideally at its full depth. Ma{\l}ek et al.\ in this volume and \cite{Krakowski} describe an independent work where the same methodology is applied to cross-matched WISE$\times$SuperCOSMOS data \citep{WISC}. In the present article we discuss the results of various tests of SVM applied to WISE-only data.

\section{Data description}
\subsection{WISE survey and our preselection}
The data used in this work come from WISE \citep{WISE}, a NASA satellite mission launched in December 2009. This 40-cm telescope, with a total $47'\times47'$ field of view, scanned the entire sky in four infrared bands ($W1$ -- $W4$) centred at 3.4, 4.6, 12 and 22 $\mu$m. 
WISE provides much better sensitivity than all the earlier infrared all-sky surveys (including IRAS, \citealt{IRAS}; 2MASS, \citealt{2MASS}; and AKARI, \citealt{AKARI}), and is free of atmospheric contamination. This directly translates to much larger photometric depth, which for WISE is about 3 mag better than for 2MASS.

The WISE catalogues are publicly available\footnote{\url{http://irsa.ipac.caltech.edu/}} and contain positional, photometric and detection quality information, as well as motion fit parameters. Here we use data from its second all-sky release, `AllWISE' \citep{AllWISE}, containing about 750 million sources, which makes it one of the largest existing astronomical catalogues. 

The goal of our work is to classify as many WISE sources as possible, therefore our basic source selection was not restrictive. In particular, we required the sources to have at least $2\sigma$ detection only in the two shortest WISE bands ($W1$ and $W2$), discarding much shallower $W3$ and $W4$ channels.
A basic cleanup to ensure reliability of the sources (removal of saturated objects and artifacts), together with an overall cut of $W1<17$ (Vega) for uniformity left us with over 606 million sources on the whole sky (see \citealt{WISC} for a map and other details). Note that owing to the $6"$ resolution of WISE, severe blending arises in the Galactic Plane and Magellanic Clouds, which makes efficient source identification practically impossible in these areas. On the other hand, WISE is only minimally affected by Galactic extinction, which is an order of magnitude smaller in the mid-IR than in the optical.

\subsection{WISE$\times$SDSS training sample}
ML source classification, such as with SVM, relies on a training sample, containing sources of already identified types.
In this work, as the training set we use the WISE data cross-matched with the spectroscopic sample from the SDSS Data Release 10 \citep{SDSS.DR10}. The resulting WISE$\times$SDSS dataset contains 2.1 million sources in total, however to ensure its reliability, we cleaned it up of insecure measurements, using in particular the SDSS \textsf{zWarning} and \textsf{zErr} parameters.
After imposing appropriate conditions on these quantities, we were left with about 390,000 stars, 1.5 million galaxies and 190,000 quasars in the training sample. 

\begin{figure*}
\begin{minipage}{0.5\textwidth}
\centering
\includegraphics[width=0.9\textwidth]{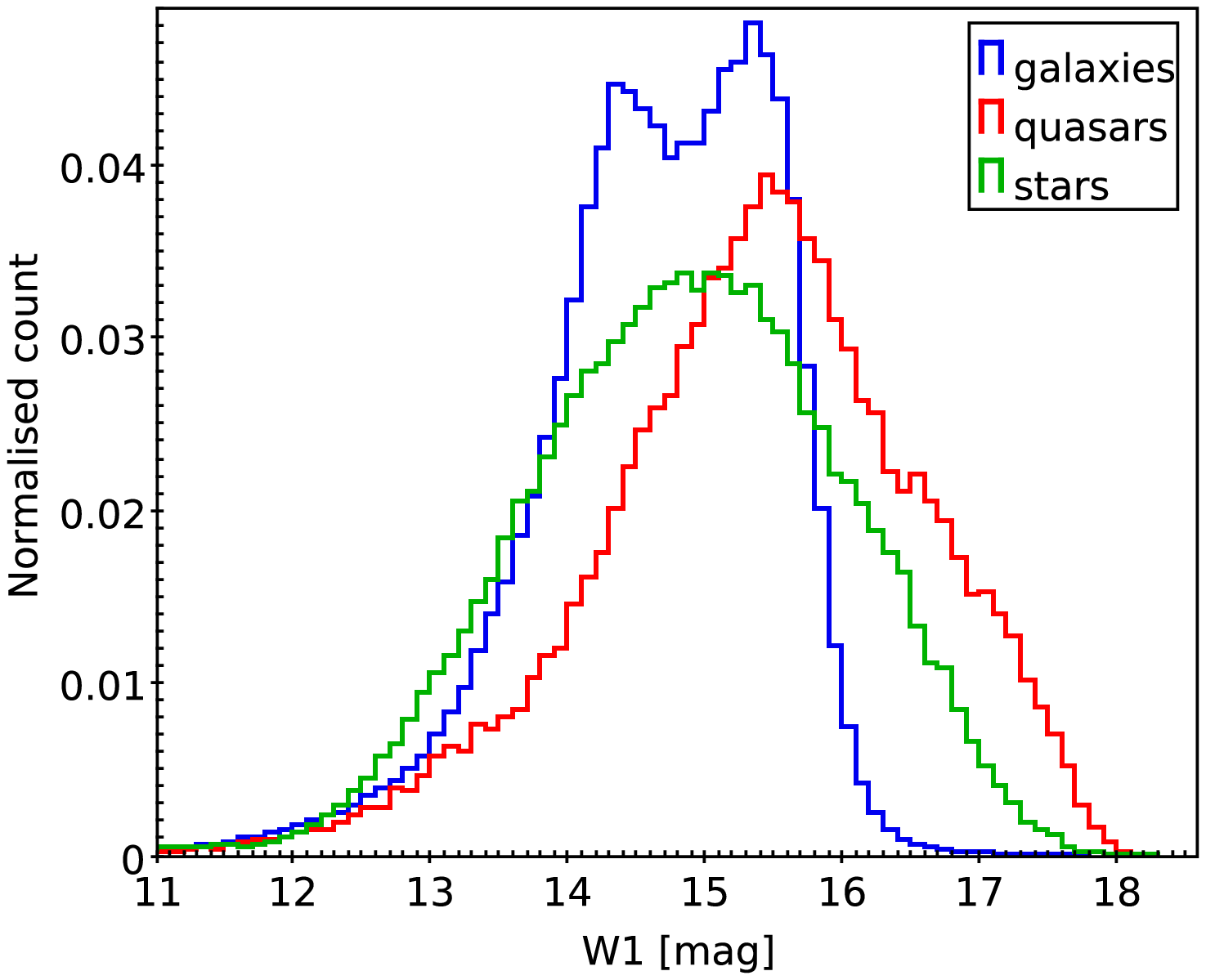}
\caption{Normalised $W1$ counts (Vega) for galaxies, quasars and stars in the WISE$\times$SDSS DR10 spectroscopic sample.}
\label{w1}
\end{minipage}
\quad
\begin{minipage}{0.5\textwidth}
\centering
\includegraphics[width=0.9\textwidth]{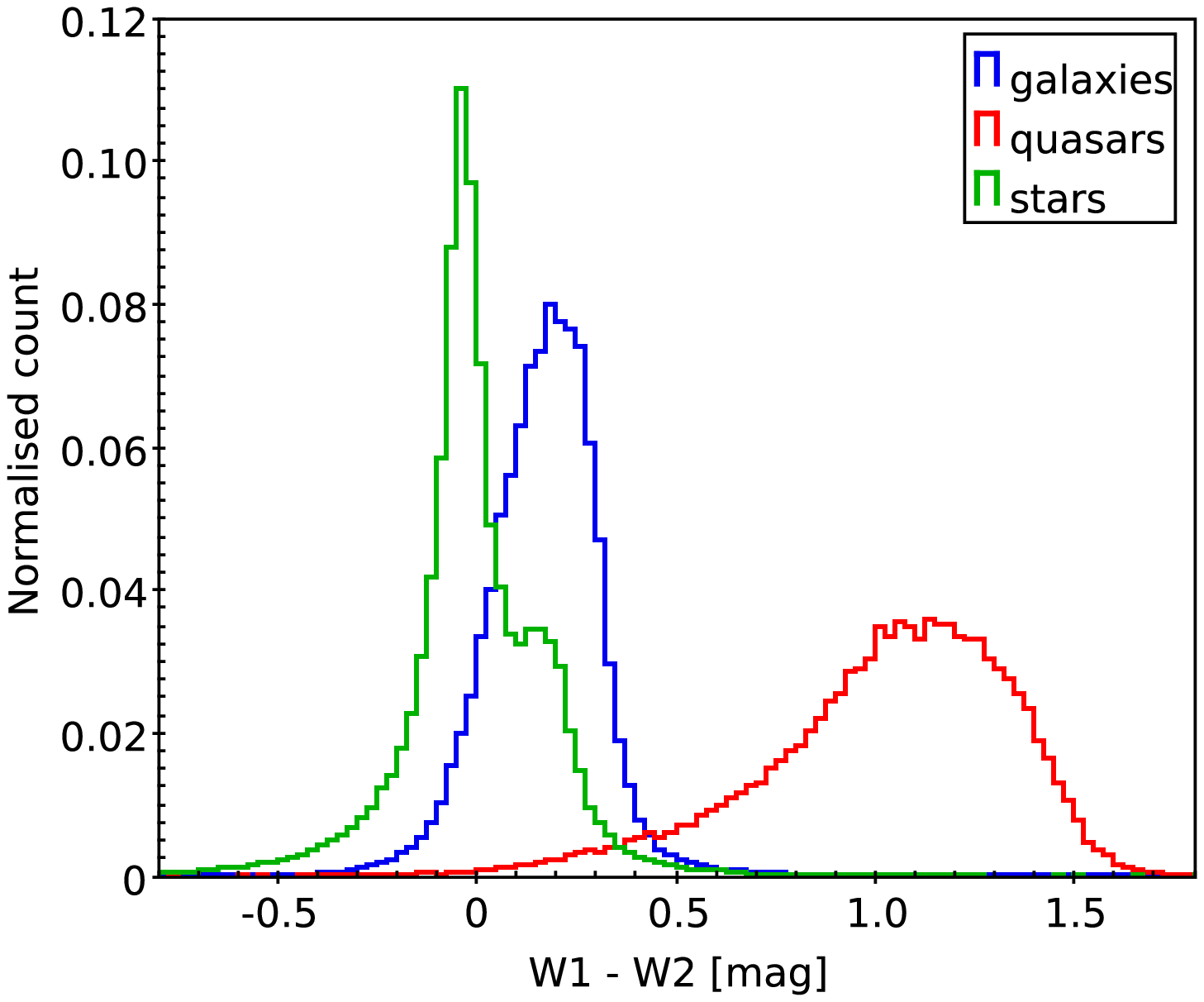}
\caption{Distribution of the $W1-W2$ colour for galaxies, quasars and stars in the WISE$\times$SDSS DR10 spectroscopic sample.}
\label{w1w2}
\end{minipage}
\end{figure*}

Figure \ref{w1} presents normalised $W1$ counts for the three source types in the WISE$\times$SDSS cross-match. As is clearly visible, SDSS contains hardly any galaxies fainter than $W1=16$. For that reason, as we will be unable to create a reliable training sample beyond that magnitude, our analysis will be from now on restricted to $W1<16$. As far as all-sky data are concerned, this cut reduces the size of our WISE catalogue to 314 million sources. As $W1=16$ is one magnitude brighter than the overall completeness of WISE, our analysis will need to be extended when deeper training samples become available, such as for instance from SDSS-IV \citep{SDSS-IV}.

We  also note that the galaxy counts in  WISE$\times$SDSS are characterized by two distinct peaks. This results from the combination of two effects: heterogeneity of the SDSS spectroscopic data, and properties of WISE-detected galaxies seen also by SDSS.
Finally, the training sample contains practically no galaxies nor quasars brighter than $W1 \sim 9.5$. This is however not a major issue in view of the eventual all-sky classification, as such WISE sources are mostly stars \citep{Jarrett11} and/or are saturated.

Figure \ref{w1w2} presents the distribution of the  $W1 - W2$ colour for the three source types. For stars and galaxies, a significant overlap in this colour makes it insufficient as a separator in the absence of other source information, such as morphology (which is \textit{not} provided in the WISE database). Even for quasars, which much more clearly separate out from other source types, the 
$W1-W2 > 0.8$ cut proposed by \cite{Stern12} is not fully appropriate, giving a potentially very incomplete sample. This further motivates our approach to identify WISE sources in multi-dimensional space  in an automatised way rather than by simple cuts.

\section{Methodology}
\subsection{SVM method of classification}
Automatic classification of the WISE sources presented here uses a class of supervised ML algorithms -- Support Vector Machines (SVMs). Similarly as some other ML classification methods, it relies on choosing an appropriate feature space, in which the sources from a given sample occupy different parts according to their class (here galaxies, stars and quasars). Thanks to an algorithm using pattern recognition -- such as SVM -- we can distinguish these classes in the multi-dimensional parameter space.

The main idea behind the SVM algorithm is to calculate a decision boundary between a set of different objects. Maximising the margin (i.e.\ the shortest distance from the decision plane to the closest points belonging to the distinct classes) between the classes closest points (the so-called support vectors), the optimal separating hyperplane between the $N$ classes of sources can be found (in our case $N=3$). For the experiments described here, we used the Gaussian radial basis kernel function, for which we tuned two parameters determining the separation boundary: $C$ and $\gamma$. The parameter $C$ is responsible for the width of the margin, while $\gamma$ specifies the topology of the decision boundary. These parameters are fitted based on the training sample, and hence the decision surface can be established.

\subsection{Efficiency of classification}

For each of the cases described hereunder, two tests were made. In the first one (\textit{self-check}), we classified the same objects as present in the training sample. In addition, for a \textit{cross-test}, we applied the classifier to a randomly chosen sample of objects outside of the training set. To verify the efficiency of our method, we computed the completeness, C, purity, P and contamination, F. For galaxies they are given by (e.g.\ \citealt{Soumagnac15}):
\begin{equation}
{\rm C_g=\rm \frac{TGG}{TGG+FGS+FGQ}}\;,
\label{Eq: completeness}
\end{equation}
\begin{equation}
\rm P_g=\rm \frac{TGG}{TGG+FSG+FQG}\;,
\end{equation}
\begin{equation}
{\rm F_g=1-P_g}\;,
\label{Eq: contamination}
\end{equation}
where TGG, FGS and FGQ refer to true galaxies classified respectively as galaxies, stars or quasars, and FSG, FQG are stars or quasars misclassified as galaxies. Similar statistics were computed for stars and quasars.

\subsection{Details of tests performed on WISE data}
\label{Sec: Details of tests}
For the classification tests presented here, we divided the WISE$\times$SDSS sample according to WISE $W1$ magnitudes and Galactic extinction to examine the behaviour of the classifier as a function of these parameters. One expects the classification efficiency to deteriorate for fainter (hence lower signal-to-noise) sources on the one hand, while extragalactic sources located in sky areas of similar extinction should have their colours similarly biased, potentially influencing the results.
We thus created three flux-limited subsamples of the training data ($W1 < 14$,  $< 15$ and $< 16$), which were further divided according to extinction.
For the latter we used the $I_{100}$ sky map, made from a combination of COBE/DIRBE and IRAS 100 $\mu$m measurements \citep{schlegel}, and applied four bins: $I_{100}\in\langle 0;1)$, $\langle 1;2)$, $\langle 2;3)$ and $\langle 3;10)$ [MJy/sr]. The training set includes practically no galaxies above $I_{100} > 10$ MJy/sr; such areas cover however mostly the Galactic plane and Magellanic Clouds where classification is unreliable anyway.

In our experiments we have used the following quantities derived from the WISE database to define the parameter space:
\begin{enumerate}
\item magnitude \textsf{w1mpro} measured with profile-fitting photometry in the $W1$ band;
\item colour $W1 - W2$ defined as the difference in the \textsf{w1mpro} and \textsf{w2mpro} profile-fitting magnitudes;
\item difference of two circular aperture magnitudes in $W1$, $\mathsf{w1mag\_1}-\mathsf{w1mag\_3}$, measured respectively in radii $5.5"$ and $11"$ ;
\item apparent motion defined as $\mathsf{pm} = \sqrt{\mathsf{pmra}^2 + \mathsf{pmdec}^2}$, where \textsf{pmra} and \textsf{pmdec} are the apparent motions in right ascension and declination, respectively.
\end{enumerate}
We note that among several dozen of photometric quantities provided in the AllWISE database, only a small fraction have reliable measurements for all the WISE sources and are not strictly correlated with each other. For that reason, the parameter space potentially available for the all-sky classification is very limited.

\section{Results}

We verified how the classifier behaves as a function of: i) the size of the training set; ii) number of training parameters; iii) Galactic extinction; iv) limiting WISE magnitude. All these tests were done in the magnitude/extinction bins. Here we briefly describe the results. More details will be provided in \cite{Kurcz16}.

\subsection{Size of the training set}

\begin{figure*}
\begin{minipage}{0.5\textwidth}
\centering
\includegraphics[width=0.9\textwidth]{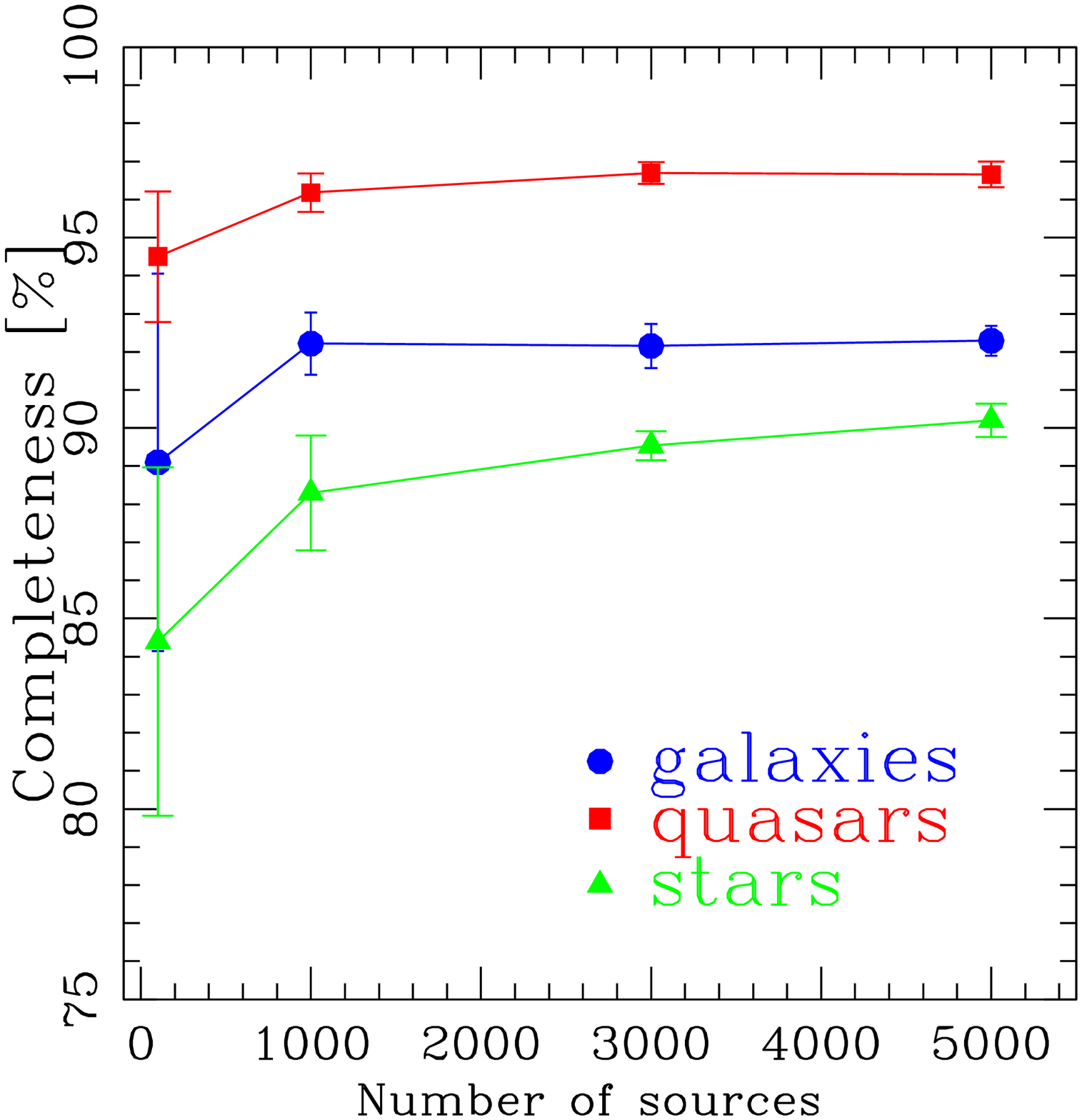}
\caption{Dependence of the completeness on the number of objects in the training set, for a bin of $W1<15$ and $I_{100}\in\langle 0;1)$ for the cross-test case.}
\label{c15}
\end{minipage}
\quad
\begin{minipage}{0.5\textwidth}
\centering
\includegraphics[width=0.9\textwidth]{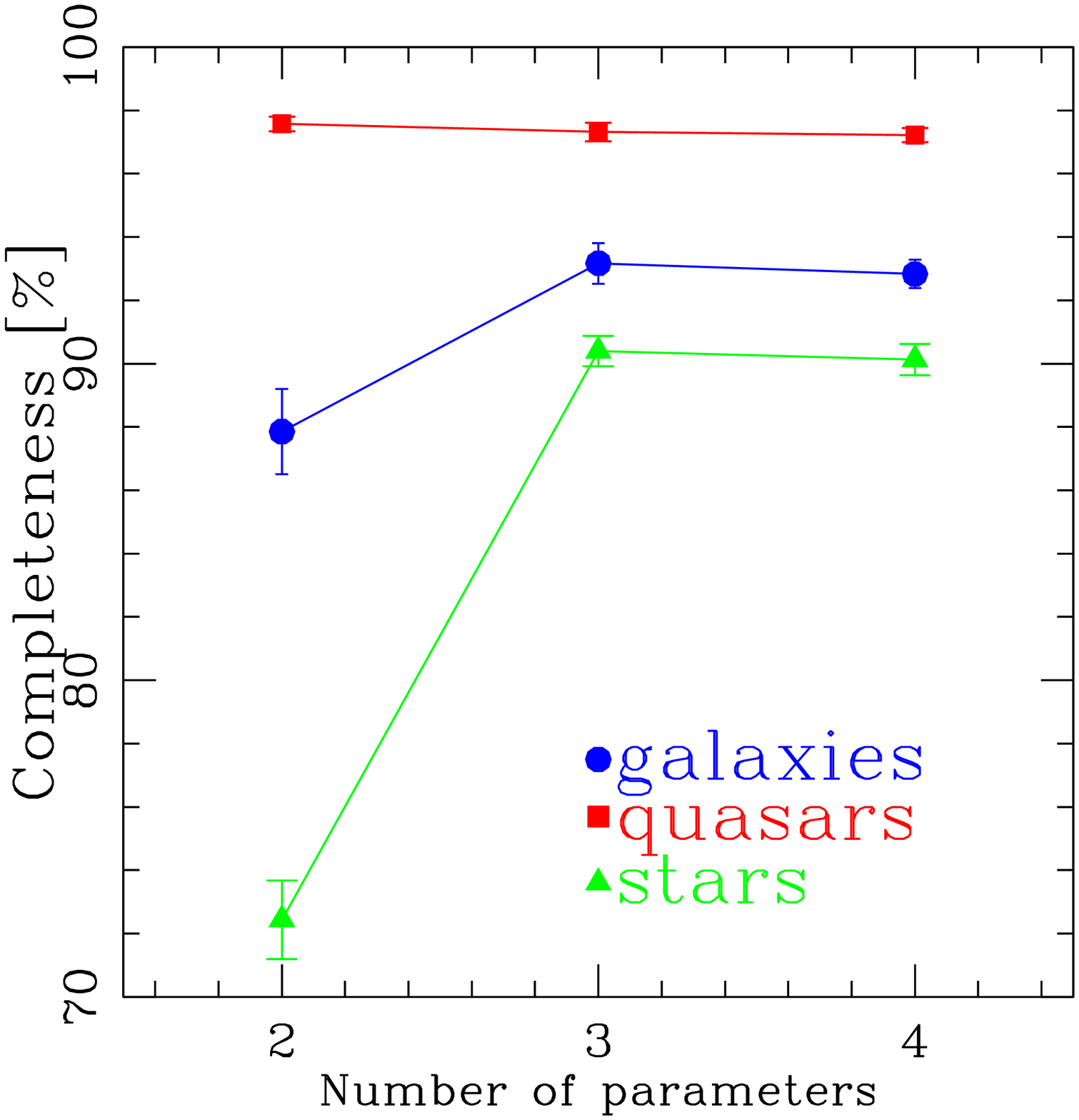}
\caption{Dependence of the completeness on the number of
classification parameters, for a bin of $W1<15$ and $I_{100}\in\langle 0;1)$ for the cross-test case.}
\label{ext05}
\end{minipage}
\end{figure*}

We first verified what is the minimum size of the training set for which the SVM classifier produces stable results. For that purpose we used random subsamples of WISE$\times$SDSS having 100, 1000, 3000 or 5000 objects of each class (i.e.\ 100 galaxies, 100 quasars and 100 stars, etc.).
Figure \ref{c15} presents an example of results: completeness for a bin $W1<15$ for the cross-test. Classifier's performance stabilises for the samples with 9000 objects in total. Similar results were obtained for the other bins, we thus conducted the remaining tests for training sets of this size.

\subsection{Dimension of the parameter space}

Having established the optimal size of the training set, we performed a series of tests to verify how many parameters would suffice for reliable classification. In this study we limited ourselves to the 4 parameters defined in Sec.\ \ref{Sec: Details of tests}.  The fourth of them, proper motions, is measured only for a fraction of WISE sources, so we used it mostly as a test-case for future, more precise measurements.

Results are presented in Figure \ref{ext05}, and the following parameter combinations are illustrated: magnitude $W1$ and colour $W1-W2$ (2 parameters); the former with the differential aperture magnitude added (3 params.); with proper motions (4 params.). Adding the differential aperture magnitude (which serves as a morphological proxy) significantly improved the results, while including also the proper motions did not. We thus conducted the remaining tests for the 3-parameter case only.

\subsection{Dependence of classification efficiency on extinction and magnitude}

\begin{figure*}
\begin{minipage}{0.5\textwidth}
\centering
\includegraphics[width=0.9\textwidth]{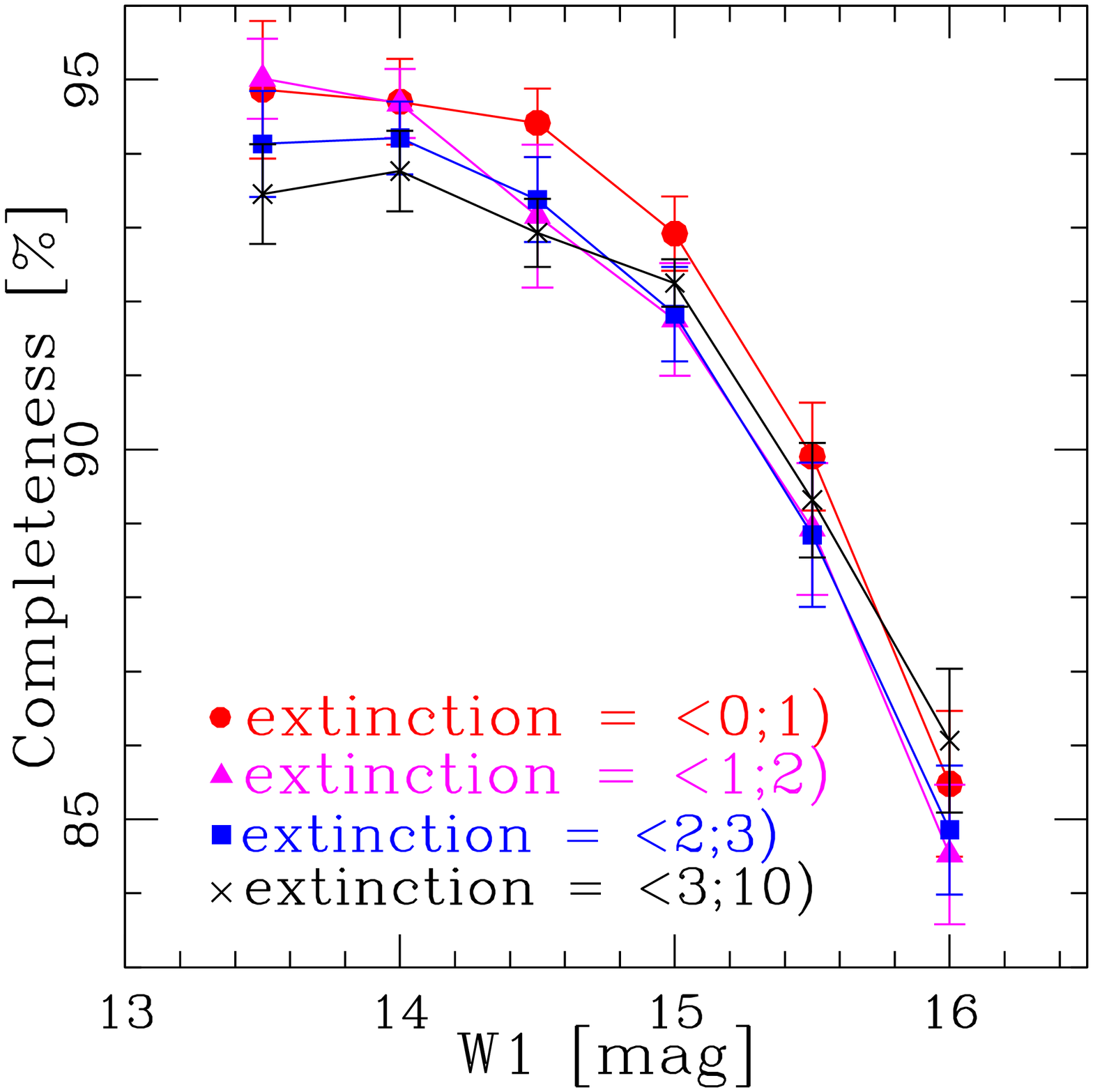}
\end{minipage}
\quad
\begin{minipage}{0.5\textwidth}
\centering
\includegraphics[width=0.9\textwidth]{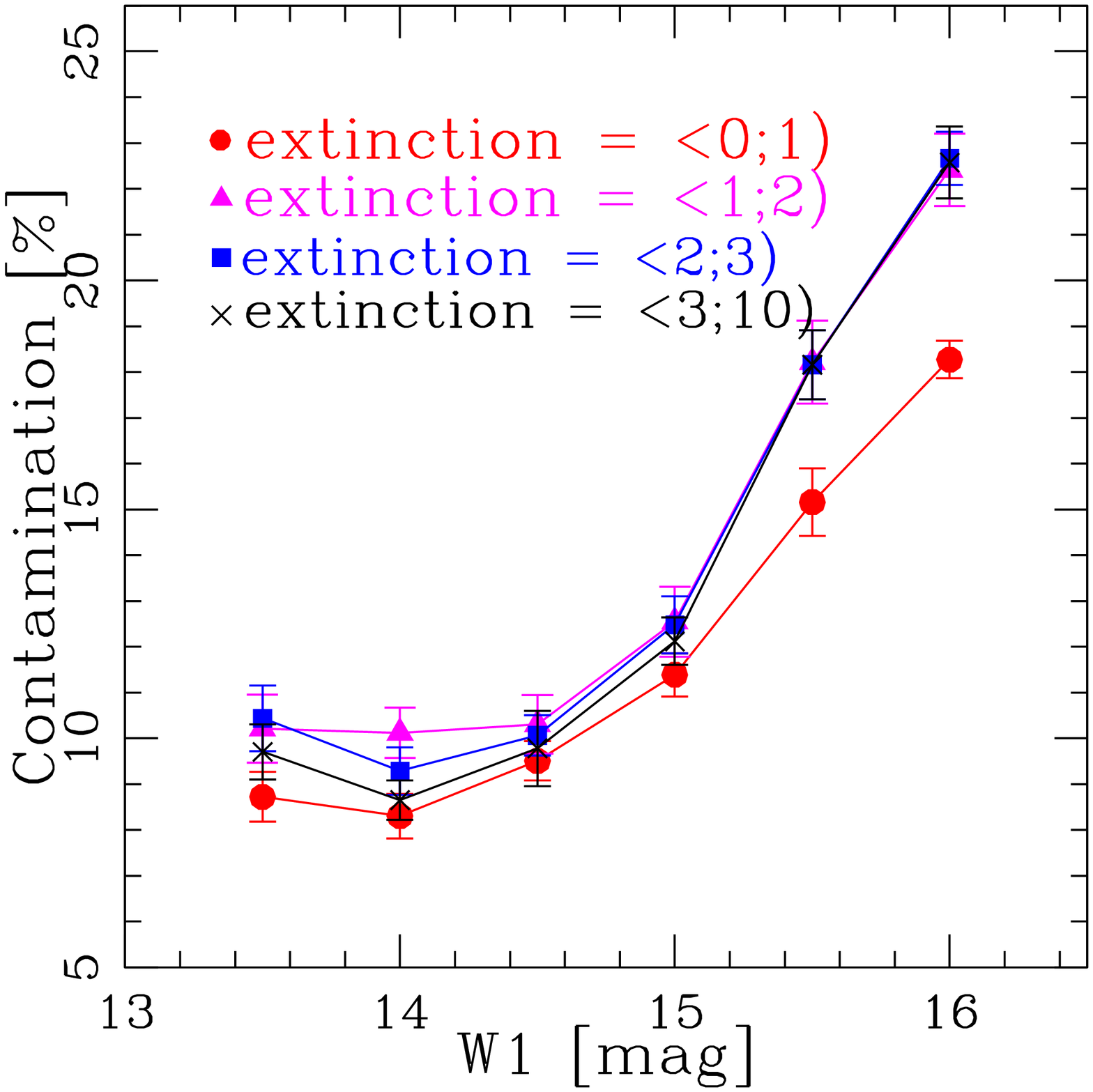}
\label{galcont}
\end{minipage}
\caption{Dependence of the completeness (left) and contamination (right) on the magnitude $W1$ for all extinction bins for galaxies. Relevant
purity levels are $100\% -$ contamination.}
\label{gal}
\quad
\begin{minipage}{0.5\textwidth}
\centering
\includegraphics[width=0.9\textwidth]{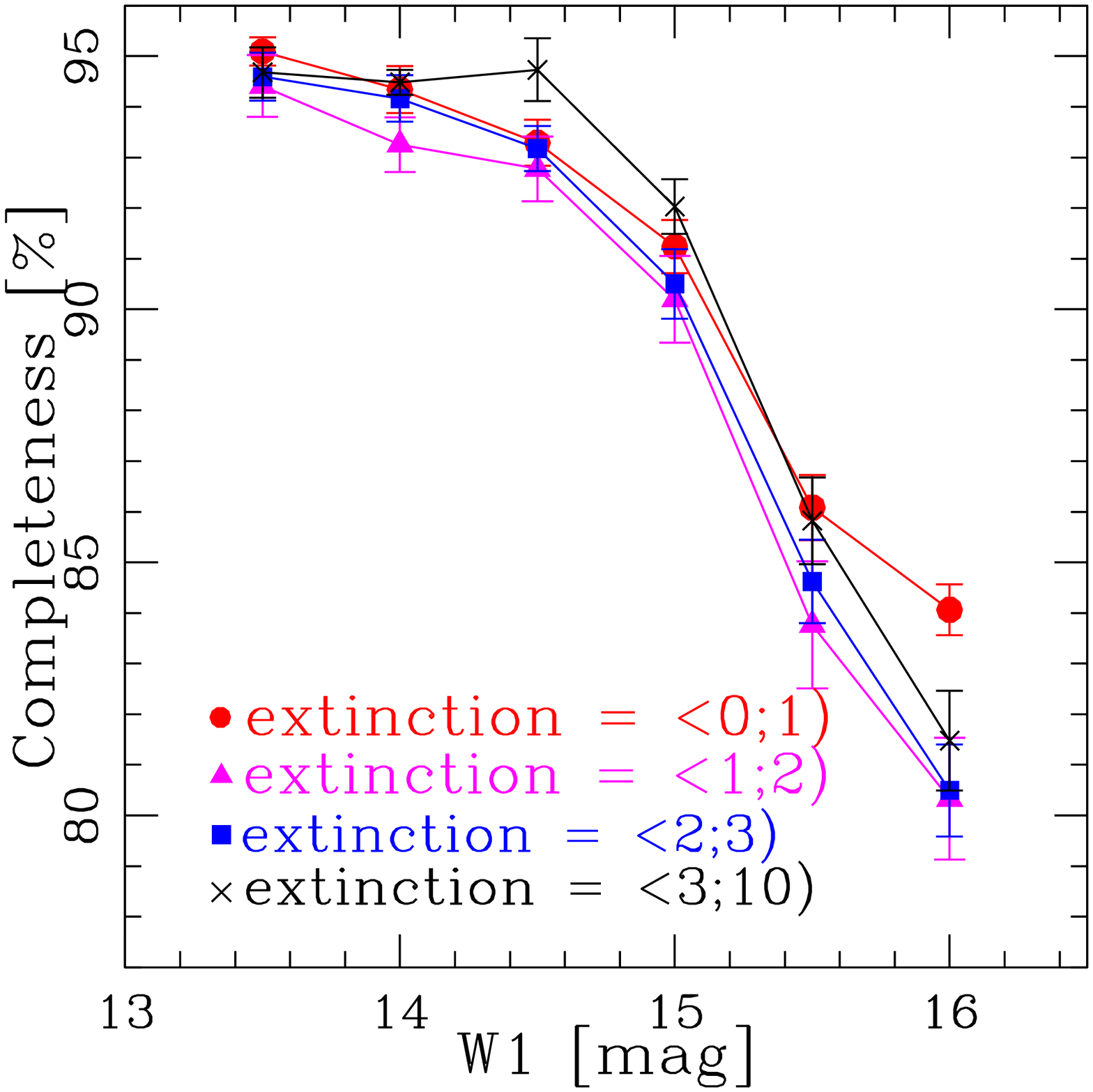}
\end{minipage}
\quad
\begin{minipage}{0.5\textwidth}
\centering
\includegraphics[width=0.9\textwidth]{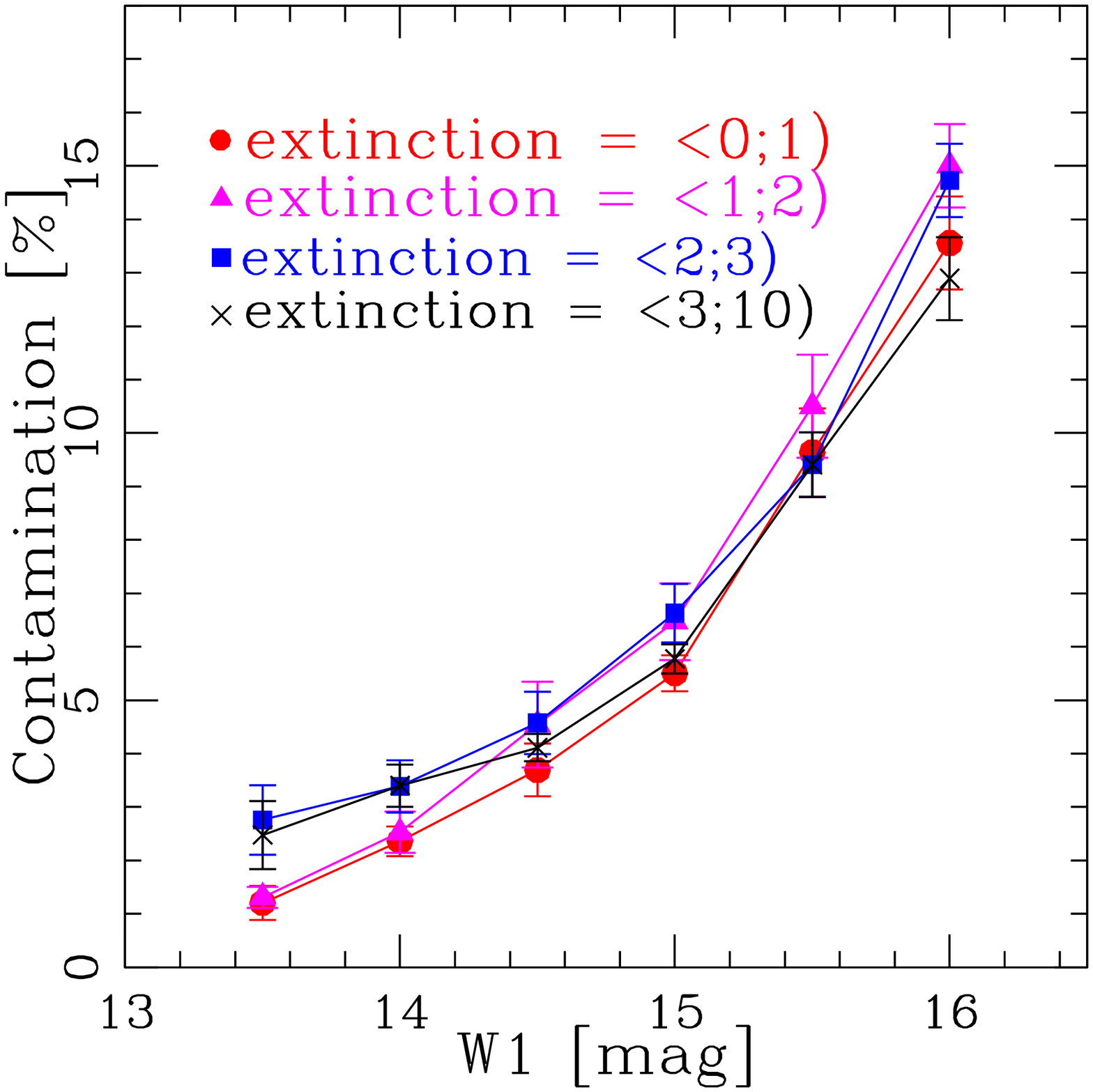}
\label{starcont}
\end{minipage}
\caption{Dependence of the completeness (left) and contamination (right) on the magnitude $W1$ for all extinction bins for stars.}
\label{star}
\quad
\begin{minipage}{0.5\textwidth}
\centering
\includegraphics[width=0.9\textwidth]{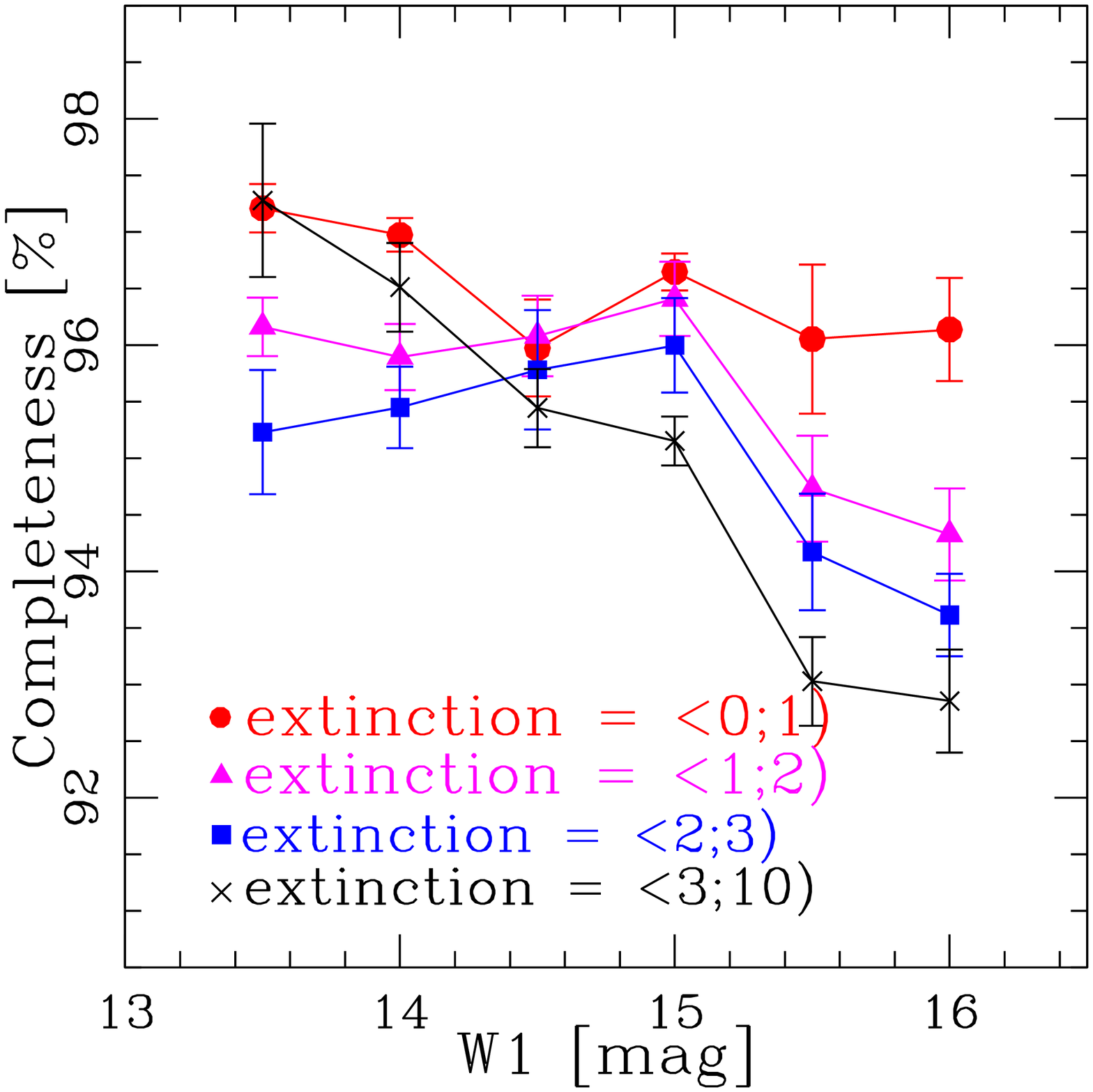}
\label{qsoo}
\end{minipage}
\quad
\begin{minipage}{0.5\textwidth}
\centering
\includegraphics[width=0.9\textwidth]{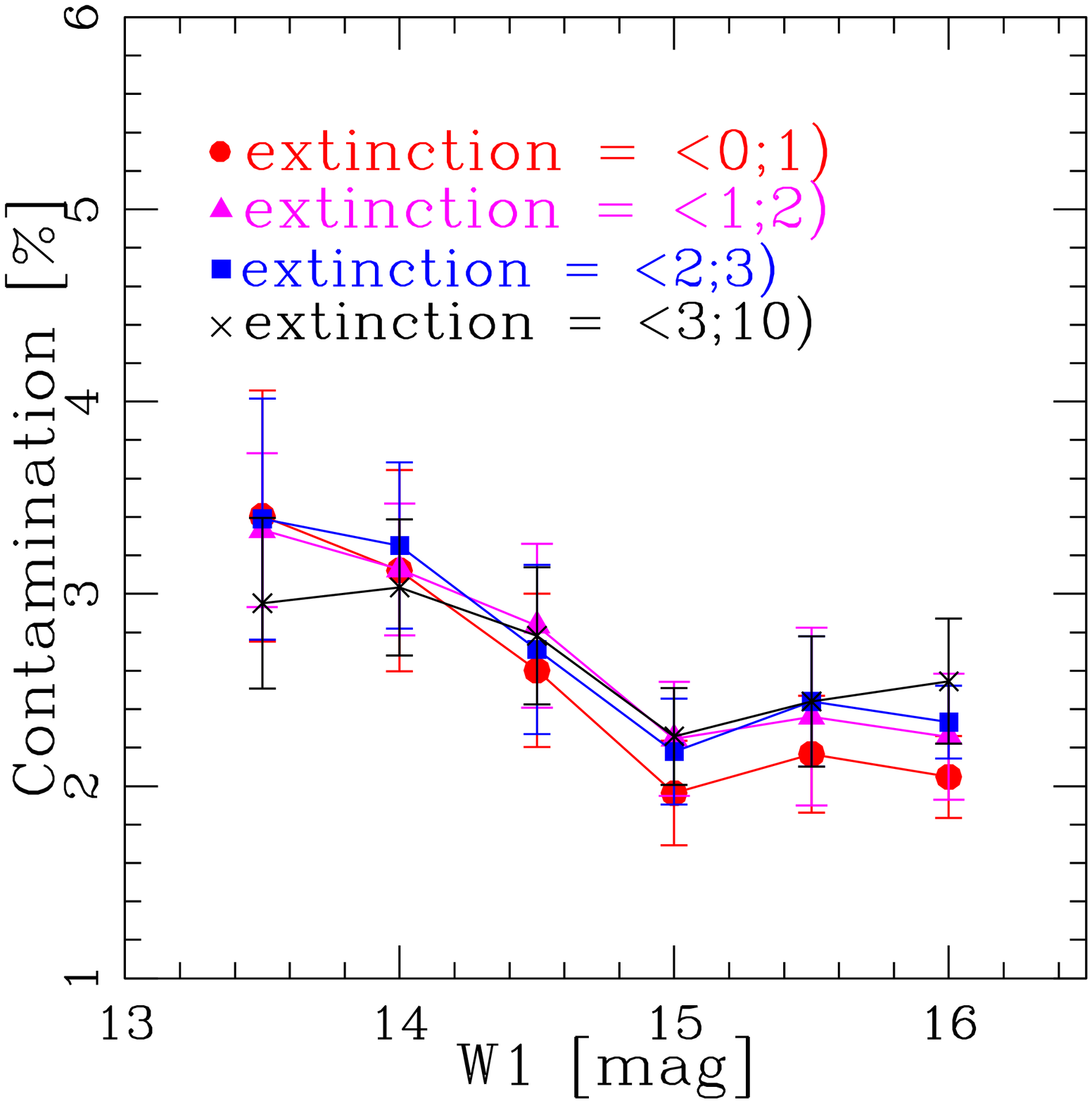}
\label{qsocont}
\end{minipage}
\caption{Dependence of the completeness (left) and contamination (right) on the magnitude $W1$ for all extinction bins for quasars.}
\label{qso}
\end{figure*}

In the final series of tests, we examined the dependence of classification statistics on the apparent $W1$ magnitude and on Galactic extinction. Here the sampling of magnitudes was done in $0.5$ mag bins, while the extinction bins remained the same as above. The results are summarised in Figures \ref{gal}--\ref{qso} for the 3 source classes. As far as the extinction is concerned, the differences between various bins are very small, consistent with no dependence of the performance on this parameter. The situation is different for magnitudes, where for galaxies and stars both the completeness and purity consistently decrease as the sources are getting fainter; there is however no such effect for quasars. Still, even at the faint end of $W1=16$, very high completeness ($\gtrsim80\%$) and relatively low contamination levels ($\lesssim25\%$) are retained for galaxies and stars. For quasars, the classifier achieved excellent performance of $\sim95\%$ completeness and $\sim3\%$ contamination for all the magnitude and extinction ranges.

\section{Future prospects}
The next step of our project will be to apply the SVM classifier, trained on the WISE$\times$SDSS data, to all-sky data from WISE. This will be presented in the forthcoming paper \citep{Kurcz16}.  Furthermore, we also plan to extend the present study by examining various SVM kernels, other ML classification methods, as well as the usability of other WISE parameters, for instance through a principal component analysis (cf.\ \citealt{Soumagnac15}). In a longer term, we plan to publicly release thus obtained galaxy, quasar and star catalogues, as we believe that reliable identification of the sources  will be of interest for the broader community.

\acknowledgements{This work was supported by the Polish National Science Center under contracts \# UMO-2012/07/D/ST9/02785 and UMO-2015/16/S/ST9/00438 (AS). AP was partially supported by the Polish-Swiss Astro Project, co-financed by a grant from Switzerland, through the Swiss Contribution to the enlarged European Union. }

\bibliographystyle{ptapap}
\bibliography{ptapapdoc}

\end{document}